\documentclass{iau}
\pdfoutput=1
\usepackage{graphicx,natbib,url}
\newcommand{\apj}{ApJ}           
\newcommand{\mnras}{MNRAS}       

\newcommand{\aap}{A\&A}
\newcommand{\aaps}{A\&AS}

\newcommand{\aj}{AJ}
\newcommand{\pasp}{PASP}
\newcommand{\apjs}{ApJS}           

\title{The Stellar Mass of M31 as inferred by the Andromeda Optical \& Infrared Disk Survey}
\shorttitle{The Stellar Mass of M31 as inferred by ANDROIDS}

\author[Sick et al]{Jonathan Sick,$^1$  Stephane Courteau,$^1$ Jean-Charles Cuillandre,$^2$ Julianne Dalcanton,$^3$ Roelof de Jong,$^4$ Michael McDonald,$^5$ Dana Simard,$^1$ \and R. Brent Tully$^6$}

\affiliation{$^1$Department of Physics, Engineering Physics \& Astronomy, Queen's University, Kingston, ON, Canada K7L 3N6. email: {\tt jsick@astro.queensu.ca}, {\tt courteau@astro.queensu.ca}\\
$^2$IRFU, Centre d'\'{e}tudes de Saclay, France {\tt jcc@cfht.hawaii.edu}\\
$^3$Department of Astronomy, University of Washington, Box 351580, Seattle, WA 98195, USA. {\tt jd@astro.washingston.edu}\\
$^4$Leibniz Institut für Astrophysik Potsdam (AIP), An der Sternwarte 16, 14482 Potsdam, Germany. {\tt rdejong@aip.de}\\
$^6$Kavli Institute for Astrophysics and Space Research, MIT, Cambridge, MA, USA. {\tt mcdonald@space.mit.edu}\\
$^6$Institute for Astronomy, University of Hawaii, 2680 Woodlawn Drive, Honolulu, HI, USA. {\tt tully@ifa.hawaii.edu}}

\pubyear{2014}
\volume{311}
\jname{Galaxy Masses as Constraints of Formation Models}
\editors{M. Cappellari \& S. Courteau, eds.}

\begin{document}

\maketitle

\begin{abstract}
Our proximity and external vantage point make M31 an ideal testbed for understanding the structure of spiral galaxies.
The Andromeda Optical and Infrared Disk Survey (ANDROIDS) has mapped M31's bulge and disk out to R=40 kpc in $ugriJK_s$ bands with CFHT using a careful sky calibration.
We use Bayesian modelling of the optical-infrared spectral energy distribution (SED) to estimate profiles of M31's stellar populations and mass along the major axis.
This analysis provides evidence for inside-out disk formation and a declining metallicity gradient.
M31's $i$-band mass-to-light ratio ($M/L_i^*$) decreases from 0.5~dex in the bulge to $\sim 0.2$~dex at 40~kpc.
The best-constrained stellar population models use the full $ugriJK_s$ SED but are also consistent with optical-only fits.
Therefore, while NIR data can be successfully modelled with modern stellar population synthesis, NIR data do not provide additional constraints in this application.
Fits to the $gi$-SED alone yield $M/L_i^*$ that are systematically lower than the full SED fit by 0.1~dex.
This is still smaller than the 0.3~dex scatter amongst different relations for $M/L_i$ via $g-i$ colour found in the literature.
We advocate a stellar mass of $M_*(30~\mathrm{kpc}) =10.3^{+2.3}_{-1.7}\times 10^{10}~\mathrm{M}_\odot$ for the M31 bulge and disk.
\end{abstract}

\firstsection
\section{Introduction}

The ANDROIDS programme has used the MegaCam and WIRCam cameras on the Canada-France-Hawaii Telescope (CFHT) to map M31's bulge and disk homogeneously within $R=40$~kpc with $ugriJK_s$ bands and enable global studies of M31's structure and stellar populations using both resolved stars and integrated spectral energy distributions (SEDs).
In this contribution, we use ANDROIDS to estimate the stellar mass profile of the M31 disk with Bayesian modelling of the optical to near-IR (NIR) SED.
This approach is more rigorous than the colour-$M/L^*$ prescriptions \citep[e.g.][]{Zibetti:2009,Taylor:2011,Into:2013} often employed by pixel-by-pixel stellar mass estimation studies that use only a $g-i$ colour and marginalize over all likely star formation histories.
By studying M31 in detail, an overall goal of ANDROIDS is to explore systematic uncertainties in studies of more distant and poorly resolved systems.

\section{M31 Surface Brightness Calibration}

Background subtraction is the most significant challenge for observational studies of M31's structure since we cannot observe its disk and blank sky simultaneously.
This is particularly acute in our NIR maps where skyglow is 3-dex brighter than the disk, while also having strong spatial and temporal variations.
In \cite{Sick:2014}, we describe our ANDROIDS/WIRCam sky-target nodding and background subtraction schemes and find that the NIR background cannot be known to better than 2\% given the scale of sky-target nods required for M31.
We overcome this uncertainty by solving for sky offsets that formally minimize surface brightness differences between overlapping pairs of images.
Such sky offsets are $\sim1$\% of the NIR brightness, but systematically uncertain up to a zeropoint normalization that is 0.16\% of the sky level.
In optical bands, the sky background is both more stable and somewhat dimmer, though we still employ sky-target nodding with the Elixir-LSB method for CFHT/MegaCam to build a real-time map of sky and scattered light backgrounds over one-hour sliding windows.
With Elixir-LSB we easily identify low surface brightness features in M31's outer disk, such as the Northern Spur, at levels below $\mu_g\sim26$~mag~arcsec$^{-2}$ \citep{Sick:2013a}.

The aforementioned sky offset zeropoint uncertainty requires that our surface brightness profiles be finely calibrated against external datasets. Resolved stellar catalogs transformed into surface brightness maps, such as our own WIRCam star catalog, and even Panchromatic Hubble Andromeda Treasury \citep[PHAT;][]{Dalcanton:2012}, provide a useful dataset up to the limit of completeness corrections. Extremely wide-field imaging is also useful as it enables a simultaneous mapping of the background and the disk light.  We are currently using Dragonfly \citep{Abraham:2014} to image M31 as a replacement for the venerable wide-field plates of \cite{Walterbos:1987}.

\section{SED Stellar Mass Modelling}

\begin{figure}
\centering
\includegraphics[width=0.7\columnwidth]{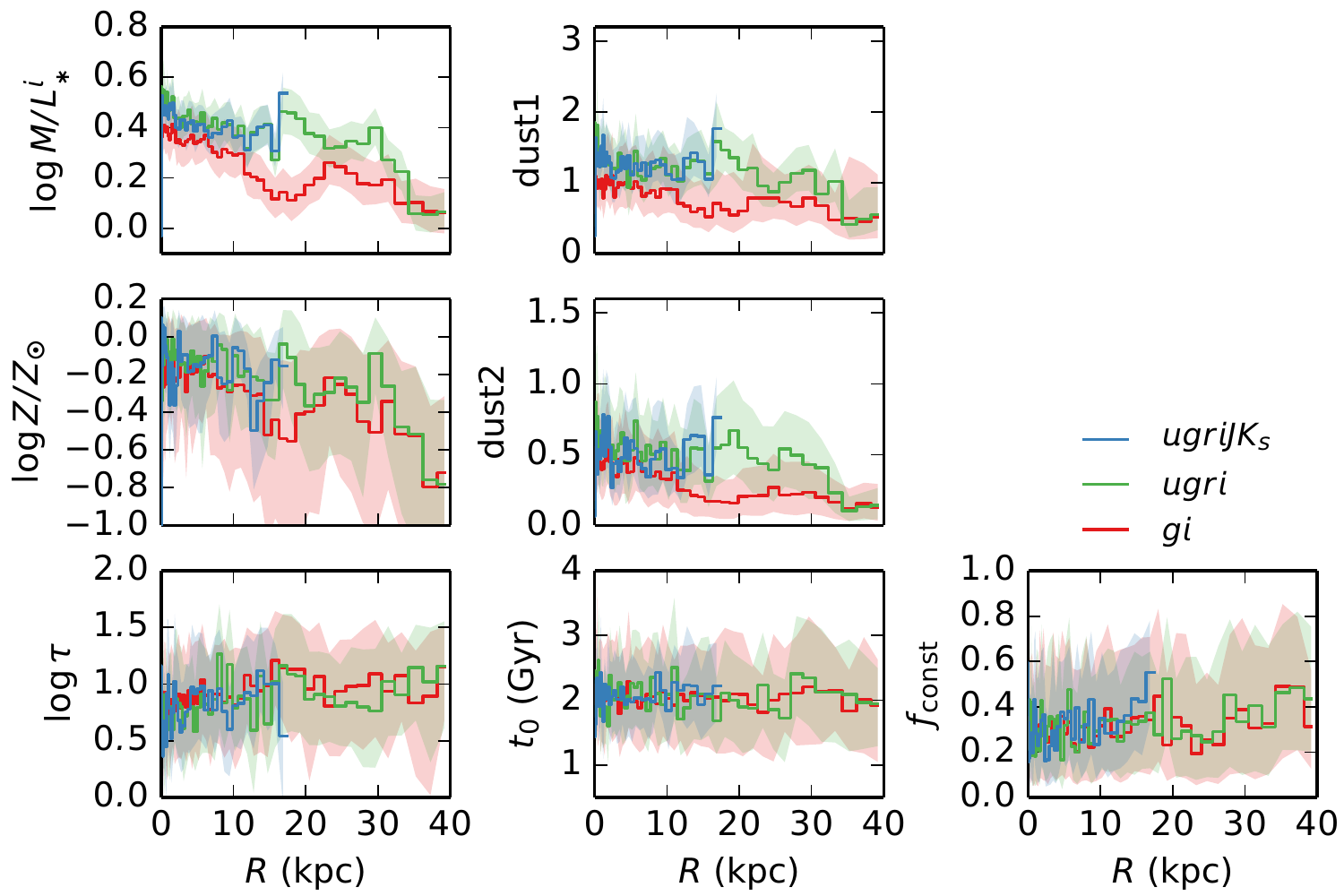} 
\caption{Posterior stellar population profiles for different bandpass sets: $ugriJK_s$ (blue), $ugri$ (green), $gi$ (red).
A declining metallicity gradient and inside-out disk formation (seen by an increase in the e-folding time, $\log \tau$, of the exponentially declining star formation history model) are clearly evident.}
\label{fig:pop_profile}
\end{figure}

\begin{figure}
\centering
\includegraphics[width=0.7\columnwidth]{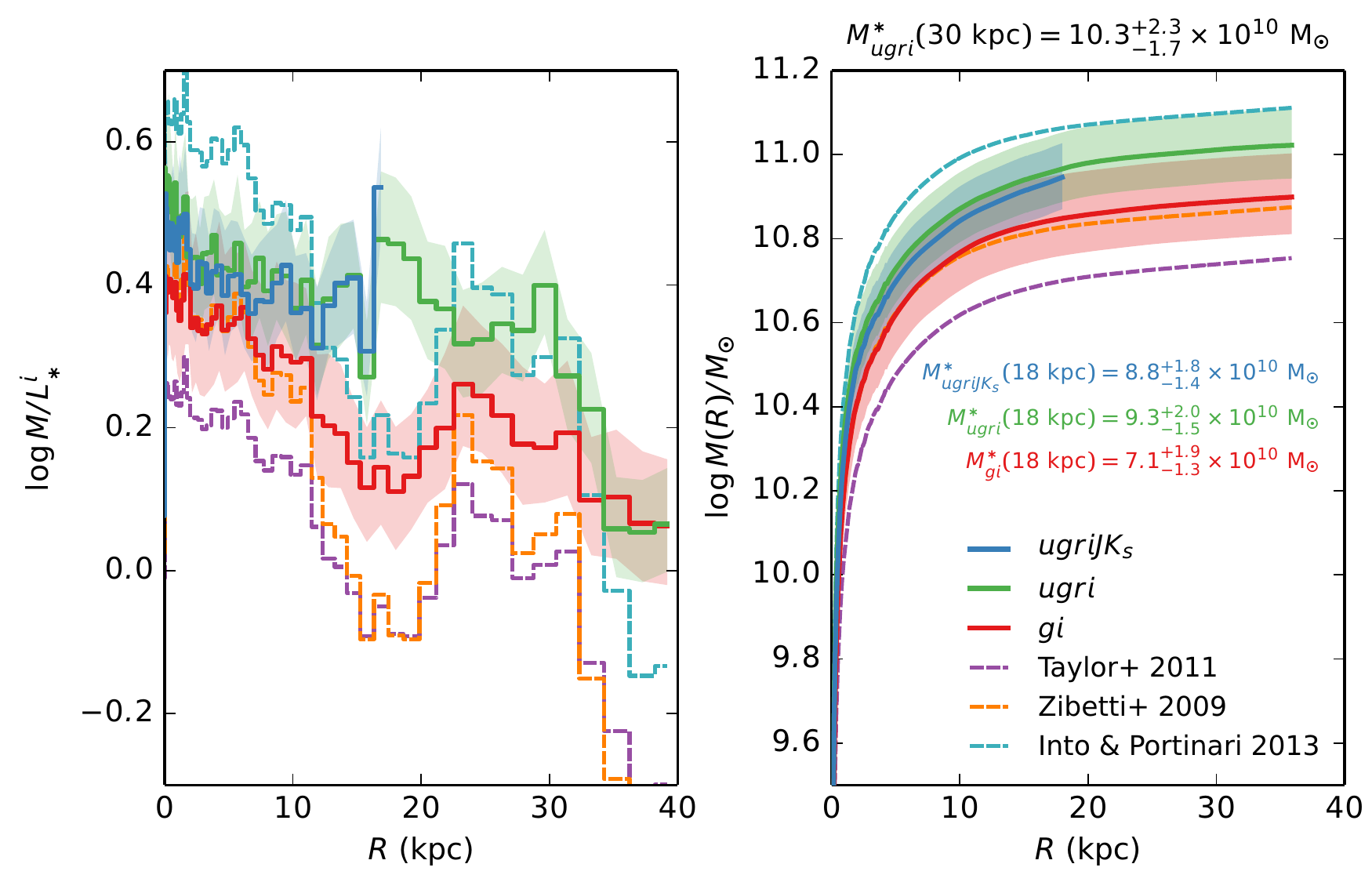} 
\caption{Posterior stellar $M/L_i^*$ (left) and stellar mass (right) major axis profiles.
The $ugriJK_s$ (blue) and $ugri$ (green) fits are consistent, while fit of only $gi$ (red) are lower by 0.1~dex in $M/L_i^*$.
Equivalent $gi$--$M/L_i^*$ relations in the literature can vary by 0.3~dex of $M/L_i^*$.}
\label{fig:mass_profile}
\end{figure}

From the calibrated surface brightness profiles, we model the SED at each radial bin to estimate the stellar population, and hence the stellar mass-to-light ratio, $M/L_i^*$.
Our modelling engine is the Flexible Stellar Population Synthesis (FSPS) software \citep{Conroy:2009,Conroy:2010}.
We chose FSPS for its reliable calibration and ``lighter'' AGB contribution than older SP models \citep[e.g.,][]{Bruzual:2003}, and allowance for deep customization of the computed stellar populations.\footnote{The lead author (J.S.) contributes to the maintenance of a Python-language wrapper for FSPS: \url{http://dan.iel.fm/python-fsps}}
We use a Markov Chain Monte Carlo approach to modelling SEDs extracted along the northern major axis of the M31 disk implemented with the \texttt{emcee} python package \citep{Foreman-Mackey:2013}.
We tested different star formation history parameterizations and found that a simple `$\tau$' model, involving constant plus exponentially declining star formation rate components minimized residuals compared to more sophisticated `delayed $\tau$' and late burst models.
Of the dust attenuation treatments, the default power-law attenuation law with separate components for young and older stellar populations also minimized residuals compared to Milky Way or starburst attenuation models.

We found that posterior SED residuals are minimized by fitting the entire $ugriJK_s$ SED.
This contradicts \cite{Taylor:2011} and \cite{Zibetti:2009} who advocated against using NIR bands in mass estimation due to uncertain AGB treatments of the previous generation of stellar population synthesis models \citep[e.g.][]{Bruzual:2003,Maraston:2005}.
Much like the NIR, the $griJK_s$-SED fit has little predictive power over the crucial $u$-band.
This result should therefore encourage SED modellers to incorporate as many bandpasses as possible, including UV and IR, to obtain the best constraints on stellar populations and masses.

We modelled SEDs extracted from a logarithmically-sized wedge \citep[e.g.][their Fig. 2]{Courteau:2011} to produce stellar population profiles (shown in Fig. \ref{fig:pop_profile}).
Interestingly, the $ugri$-fit and $ugriJK_s$-fit SEDs produce statistically identical stellar population profiles, with the only exception being a slightly tighter posterior credible region from the full-SED fits.
Although the consistency of optical and optical-NIR SED fits is reassuring from the perspective of NIR calibrations, it is also disappointing that the NIR data has not produced a remarkably improved posterior stellar population estimate.

Clearly evident is that poorly sampled SEDs can bias results.
Fitting only the $gi$ SED (that is, using an input information equivalent to those using colour-$M/L^*$ look-up-tables) clearly biases the posterior stellar population distribution, with significantly lower dust opacities and lower mass-to-light ratios.
By comparison, we have also plotted mass-to-light ratios predicted by three colour-$M/L^*$ relations \citep{Zibetti:2009,Taylor:2011,Into:2013}.
These fits systematically vary by 0.3~dex, far larger than the 0.1~dex of internal systematic uncertainty typically claimed by $g-i$ -- $M/L^*$ fits \cite{Courteau:2013}.
Compared to our full SED fits, modelling of the $gi$ SED is less biased than these other $M/L^*$ fits, which are based on other stellar population synthesis models.
This serves as reminder that stellar mass estimates remain dominated by prior assumptions such as choices of IMF, dust, and details of AGB treatments, among other concerns.

\section{Discussion}

We have used $ugriJK_s$ SEDs to map the stellar mass of M31's disk and find a stellar mass, within $30$~kpc, of $M_{ugri}^{*} = 10.3^{+2.3}_{-1.7}\times 10^{10}~\mathrm{M}_\odot$.
This result is consistent with the stellar bulge and disk masses quoted by \cite{Tamm:2012} ($10.1\times10^{10}~\mathrm{M}_\odot$).
Future work will extend this analysis to a full 2D mapping of the M31 stellar mass distribution.

We are matching these stellar mass maps with dynamical tracers of gas and stars to construct a mass model of M31's stellar, gas, and dark matter components (Simard et~al., in progress). 
The DiskFit code \citep{Spekkens:2007} allows us to correct the H\textsc{i} velocity fields of \cite{Saglia:2010} and \cite{Chemin:2009} for non-circular motions in M31's central parts. 
The success of this mass model will be determined by the stability of a dynamical N-body realization.

\section*{Acknowledgements}

\noindent J.S. and S.C. acknowledge support through respective Graduate Scholarship and Discovery grants from the Natural Sciences and Engineering Research Council of Canada. We thank the Canadian Advanced Network for Astronomical Research (CANFAR) for enabling the computing facilities needed for this work.


\begin{thebibliography}{19}
\expandafter\ifx\csname natexlab\endcsname\relax\def\natexlab#1{#1}\fi

\bibitem[{{Abraham} \& {van Dokkum}(2014)}]{Abraham:2014}
{Abraham}, R.~G., \& {van Dokkum}, P.~G. 2014, \pasp, 126, 55

\bibitem[{{Bruzual} \& {Charlot}(2003)}]{Bruzual:2003}
{Bruzual}, G., \& {Charlot}, S. 2003, \mnras, 344, 1000

\bibitem[{{Chemin} {et~al.}(2009){Chemin}, {Carignan}, \&
  {Foster}}]{Chemin:2009}
{Chemin}, L., {Carignan}, C., \& {Foster}, T. 2009, \apj, 705, 1395

\bibitem[{{Conroy} {et~al.}(2009){Conroy}, {Gunn}, \& {White}}]{Conroy:2009}
{Conroy}, C., {Gunn}, J.~E., \& {White}, M. 2009, \apj, 699, 486

\bibitem[{{Conroy} {et~al.}(2010){Conroy}, {White}, \& {Gunn}}]{Conroy:2010}
{Conroy}, C., {White}, M., \& {Gunn}, J.~E. 2010, \apj, 708, 58

\bibitem[{{Courteau} {et~al.}(2011){Courteau}, {Widrow}, {McDonald},
  {Guhathakurta}, {Gilbert}, {Zhu}, {Beaton}, \& {Majewski}}]{Courteau:2011}
{Courteau}, S., {Widrow}, L.~M., {McDonald}, M., {et~al.} 2011, \apj, 739, 20

\bibitem[{{Courteau} {et~al.}(2014){Courteau}, {Cappellari}, {de Jong},
  {Dutton}, {Emsellem}, {Hoekstra}, {Koopmans}, {Mamon}, {Maraston}, {Treu}, \&
  {Widrow}}]{Courteau:2013}
{Courteau}, S., {Cappellari}, M., {de Jong}, R.~S., {et~al.} 2014, Rev. Mod.
  Phys., 86, 47

\bibitem[{{Dalcanton} {et~al.}(2012){Dalcanton}, {Williams}, {Lang}, {Lauer},
  {Kalirai}, {Seth}, {Dolphin}, {Rosenfield}, {Weisz}, {Bell}, {Bianchi},
  {Boyer}, {Caldwell}, {Dong}, {Dorman}, {Gilbert}, {Girardi}, {Gogarten},
  {Gordon}, {Guhathakurta}, {Hodge}, {Holtzman}, {Johnson}, {Larsen}, {Lewis},
  {Melbourne}, {Olsen}, {Rix}, {Rosema}, {Saha}, {Sarajedini}, {Skillman}, \&
  {Stanek}}]{Dalcanton:2012}
{Dalcanton}, J.~J., {Williams}, B.~F., {Lang}, D., {et~al.} 2012, \apjs, 200,
  18

\bibitem[{{Foreman-Mackey} {et~al.}(2013){Foreman-Mackey}, {Hogg}, {Lang}, \&
  {Goodman}}]{Foreman-Mackey:2013}
{Foreman-Mackey}, D., {Hogg}, D.~W., {Lang}, D., \& {Goodman}, J. 2013, \pasp,
  125, 306

\bibitem[{{Into} \& {Portinari}(2013)}]{Into:2013}
{Into}, T., \& {Portinari}, L. 2013, \mnras, 430, 2715

\bibitem[{{Maraston}(2005)}]{Maraston:2005}
{Maraston}, C. 2005, \mnras, 362, 799

\bibitem[{{Saglia} {et~al.}(2010){Saglia}, {Fabricius}, {Bender}, {Montalto},
  {Lee}, {Riffeser}, {Seitz}, {Morganti}, {Gerhard}, \& {Hopp}}]{Saglia:2010}
{Saglia}, R.~P., {Fabricius}, M., {Bender}, R., {et~al.} 2010, \aap, 509, A61

\bibitem[{{Sick} {et~al.}(2013){Sick}, {Courteau}, \&
  {Cuillandre}}]{Sick:2013a}
{Sick}, J., {Courteau}, S., \& {Cuillandre}, J.-C. 2013, 1310.4832

\bibitem[{{Sick} {et~al.}(2014){Sick}, {Courteau}, {Cuillandre}, {McDonald},
  {de Jong}, \& {Tully}}]{Sick:2014}
{Sick}, J., {Courteau}, S., {Cuillandre}, J.-C., {et~al.} 2014, \aj, 147, 109

\bibitem[{{Spekkens} \& {Sellwood}(2007)}]{Spekkens:2007}
{Spekkens}, K., \& {Sellwood}, J.~A. 2007, \apj, 664, 204

\bibitem[{{Tamm} {et~al.}(2012){Tamm}, {Tempel}, {Tenjes}, {Tihhonova}, \&
  {Tuvikene}}]{Tamm:2012}
{Tamm}, A., {Tempel}, E., {Tenjes}, P., {Tihhonova}, O., \& {Tuvikene}, T.
  2012, \aap, 546, A4

\bibitem[{{Taylor} {et~al.}(2011){Taylor}, {Hopkins}, {Baldry}, {Brown},
  {Driver}, {Kelvin}, {Hill}, {Robotham}, {Bland-Hawthorn}, {Jones}, {Sharp},
  {Thomas}, {Liske}, {Loveday}, {Norberg}, {Peacock}, {Bamford}, {Brough},
  {Colless}, {Cameron}, {Conselice}, {Croom}, {Frenk}, {Gunawardhana},
  {Kuijken}, {Nichol}, {Parkinson}, {Phillipps}, {Pimbblet}, {Popescu},
  {Prescott}, {Sutherland}, {Tuffs}, {van Kampen}, \&
  {Wijesinghe}}]{Taylor:2011}
{Taylor}, E.~N., {Hopkins}, A.~M., {Baldry}, I.~K., {et~al.} 2011, \mnras, 418,
  1587

\bibitem[{{Walterbos} \& {Kennicutt}(1987)}]{Walterbos:1987}
{Walterbos}, R.~A.~M., \& {Kennicutt}, Jr., R.~C. 1987, \aaps, 69, 311

\bibitem[{{Zibetti} {et~al.}(2009){Zibetti}, {Charlot}, \&
  {Rix}}]{Zibetti:2009}
{Zibetti}, S., {Charlot}, S., \& {Rix}, H. 2009, \mnras, 400, 1181

\end{thebibliography}
\end{document}